\documentstyle[11pt,newpasp,twoside]{article}
\def\edcomment#1{\iffalse\marginpar{\raggedright\sl#1\/}\else\relax\fi}
\reversemarginpar

\begin{document}

\title{Portent of Heine's Reciprocal Square Root Identity}
\author{Howard S. Cohl\altaffilmark{1}}
\affil{Lawrence Livermore National Laboratory, 7000 East Ave, \\
       Livermore, CA, 94550, U.S.A.}

\altaffiltext{1}{Currently at School of Physics, University of Exeter, UK}

\begin{abstract}
Precise efforts in theoretical astrophysics are needed to fully understand 
the mechanisms that govern the structure, stability, dynamics, formation, 
and evolution of differentially rotating stars.  Direct computation of the 
physical attributes of a star can be facilitated by the use of highly 
compact azimuthal and separation angle Fourier formulations of the Green's
functions for the linear partial differential equations of mathematical physics.  
\end{abstract}

\section{Introduction}

Solutions to the three dimensional inhomogeneous linear partial differential 
equations of mathematical physics are expressible in terms their appropriate 
Green's functions (Duffy 2001).  Single, double, and triple eigenfunction 
integration and summation expressions for Green's functions are obtainable 
in the coordinate systems that allow for separation of variables (Morse \& 
Feshbach 1953).  Separation of variables maps the linear three dimensional 
homogeneous partial differential equations into three decoupled ordinary 
differential equations (Miller 1977).  Important linear 3D partial differential 
equations of mathematical physics include the Laplace equation, the Helmholtz 
equation, the biharmonic equation, the wave equation, the diffusion equation, 
and the Schroedinger equation.  The properties of R-separability for these 
partial differential equations are reflected in the Green's function 
expansions.  In the rotationally invariant coordinate systems that R-separably 
solve these equations, new discrete Fourier representations exist for the 
Green's functions for the Laplace equation (Cohl et al. 2000).  Discrete
Fourier expansions, given in terms of the azimuthal and separation angles,
must exist for the rest of the linear partial differential equations of 
mathematical physics.  The coefficients of these discrete Fourier 
representations will be given in terms of identifiable transcendental 
functions  obtained by reversing and collapsing Green's function expansions 
for these partial differential equations.  The fundamental mathematical 
tools required in order to complete this investigation are commonly available 
in the mathematics and physics literature.  

The Green's function for Laplace's equation is
\begin{equation}
{\mathcal{G}}_L({\bf x},{\bf x}') = \frac{1}{|{\bf x}-{\bf x}'|},
\end{equation}

\noindent where ${\bf x}$ and ${\bf x}'$ are given by the Cartesian
coordinates $(x,y,z)$ and $(x',y',z')$.  Traditionally, the Green's
function is expanded in terms of spherical harmonics
\begin{equation}
{\mathcal{G}}_L = \frac{1}{\sqrt{rr'}}\sum_{\ell=0}^\infty \biggl( \frac{r_<}{r_>}\biggr)^{\ell+\frac{1}{2}}
P_\ell(\cos\gamma),
\end{equation}
\noindent in spherical $(r,\phi,\theta)$ coordinates, where 
$r_<$ ($r_>$) is the smaller (larger) of the distances $r$ and $r'$,
$\cos\gamma = \cos\theta\cos\theta'+ \sin\theta\sin\theta'\cos(\phi-\phi'),$
and $P_\ell$ is the Legendre polynomial of the first kind. 
\footnote{``...The set of six coordinates $\{{\bf x},{\bf x'}\}$ may
be viewed either as defining two points relative to the origin
or as the coordinates of a three-body system once the center of
mass has been separated.  In the {\em body frame,} three out
of the six coordinates are dynamical variables, the potential
energy depending on them.'' (Cohl et al. 2001)}

By utilizing Heine's toroidal 
identity (Cohl \& Tohline 1999, Cohl et al. 2000, Cohl et al. 2001):
\begin{equation}
\frac{1}{\sqrt{\zeta-\cos\psi}} = 
\frac{\sqrt{2}}{\pi} \sum_{m=-\infty}^\infty 
Q_{m-\frac{1}{2}}(\zeta)\  e^{im\psi},
\end{equation}
\noindent where $|\zeta|\ge 1,$ and, $0\le\psi\le 2\pi,$ one can derive 
two highly compact Fourier series representations for the Green's 
function of Laplace's equation
\begin{equation}
{\mathcal{G}}_L = \frac{1}{\pi\sqrt{rr'}}\sum_{n=-\infty}^\infty
Q_{n-\frac{1}{2}}\biggl(\frac{r^2+r'^2}{2rr'}\biggr) e^{in\gamma},
\end{equation}

\begin{equation}
{\mathcal{G}}_L = \frac{1}{\pi\sqrt{RR'}}\sum_{m=-\infty}^\infty
Q_{m-\frac{1}{2}}\biggl(\frac{R^2+R'^2+(z-z')^2}{2RR'}\biggr) e^{im(\phi-\phi')},
\end{equation}

\noindent in spherical and cylindrical $(R,\phi,z)$ coordinates respectively.
Similar discrete Fourier transform expressions must exist for the linear 
partial differential equations of mathematical physics.  Algorithmic implementation of compact 
Fourier representations for the Green's functions of mathematical physics 
significantly improves the accuracy of the solution for these partial differential equations.  

\section{The Biharmonic, Triharmonic, and Higher Harmonic Equations}

One may also express potential problems in terms of an inhomogeneous 
biharmonic equation whose source is proportional to the Laplacian
of the density.  The solution is expressible in terms of an integral 
of it's Green's function, $|{\bf x}-{\bf x}'|$, convolved with source.
The resulting inhomogeneous biharmonic equation
\begin{equation}
\nabla^4 \Phi({\bf x}) = 4\pi\ \nabla^2 \rho({\bf x}),
\end{equation}

\noindent where $\nabla^2$ is the Laplacian, represents the solution 
to the potential problem over the infinite three-dimensional domain.  
The Green's function for the biharmonic equation (Vautherin 1972)
\begin{equation}
{\mathcal{G}}_{2L}({\bf x},{\bf x}') = \frac{1}{2}|{\bf x}-{\bf x}'|,
\end{equation}

\noindent is a much better behaved integration kernel than for Laplace's
equation because the singularity has been moved to infinity through 
\begin{equation}
\Phi({\bf x})=-\frac{1}{2}\int\  d^3{\bf x'}\  \nabla'\ ^2 \rho({\bf x'})\ |{\bf x}-{\bf x'}|.
\end{equation}
One can
generalize Heine's reciprocal square root identity using
(Abramowitz \& Stegun, 1965; Gradshteyn \& Ryzhik 1994--to within a sign)

\begin{equation}
\int_\zeta^\infty\cdots\int_\zeta^\infty Q_{m-\frac{1}{2}}(\zeta)(d\zeta)^n=
(-1)^n
\frac{\Gamma(m-n+\frac{1}{2})}{\Gamma(m+n+\frac{1}{2})}
(\zeta^2-1)^{n/2} Q_{m-\frac{1}{2}}^n(\zeta),
\end{equation}

\noindent and by integrating both sides of eq. (2) with respect to $\zeta.$ 
%% Consequently, the following can be shown,
%% \begin{equation}
%% \frac{d^nQ_{m-\frac{1}{2}}(\zeta)}{d\zeta^n} =
%% \frac{\Gamma(m+n+\frac{1}{2})}{\Gamma(m-n+\frac{1}{2})}
%% \frac{(-1)^n}{(\zeta^2-1)^n}\int_\zeta^\infty\cdots\int_\zeta^\infty
%% (dz)^n Q_{m-\frac{1}{2}}(\zeta),
%% \end{equation}
%% 
%% \noindent and,
%% 
%% \begin{equation}
%% \int_\zeta^\infty\cdots\int_\zeta^\infty (dz)^n Q_{m-\frac{1}{2}}(\zeta)=
%% (-1)^n\frac{\Gamma(m-n+\frac{1}{2})}{\Gamma(m+n+\frac{1}{2})}
%% \ (\zeta^2-1)^{\frac{n}{2}} Q_{m-\frac{1}{2}}^n(\zeta).
%% nd{equation}
Performing the integrations over $\zeta$ we obtain the following general 
result for {\it any} integer $n$ - generalizing the last expression at the 
bottom of p. 182 in Magnus, Oberhettinger, \& Soni (1966)
(see also eqs. (26) \& (30) in Cohl et al. 2000)
\begin{equation}
(\zeta-\cos\psi)^{n-\frac{1}{2}} = 
\sqrt{\frac{2}{\pi}} \frac{(\zeta^2-1)^{\frac{n}{2}}}{\Gamma(\frac{1}{2}-n)}  \sum_{m=-\infty}^\infty 
\frac{\Gamma(m-n+\frac{1}{2})}{\Gamma(m+n+\frac{1}{2})}\ 
Q_{m- \frac{1}{2} }^n(\zeta)\ e^{im\psi},
\end{equation}

\noindent which reduces for $n=1$ to

\begin{equation}
\sqrt{\zeta-\cos\psi} = 
\frac{\sqrt{\zeta^2-1}}{\sqrt{2}\pi} \sum_{m=-\infty}^\infty 
\biggl(m^2-\frac{1}{4}\biggr)^{-1}
Q_{m-\frac{1}{2}}^1(\zeta)\ e^{im\psi}.
\end{equation}

\noindent Using this expression, we can rewrite the Green's function
for the biharmonic equation in two different ways

\begin{equation}
2{\mathcal{G}}_{2L} = |{\bf x}-{\bf x}'| = \frac{r_>^2-r_<^2}{2\pi\sqrt{rr'}}\sum_{n=-\infty}^\infty
\biggl(n^2-\frac{1}{4}\biggr)^{-1}
Q_{n-\frac{1}{2}}^1\biggl(\frac{r^2+r'^2}{2rr'}\biggr) e^{in\gamma},
\end{equation}

\noindent and

\begin{equation}
2{\mathcal{G}}_{2L} = 
\Psi
\sum_{m=-\infty}^\infty
\biggl(m^2 - \frac{1}{4}\biggr)^{-1}
Q_{m-\frac{1}{2}}^1\biggl(\frac{R^2+R'^2+(z-z')^2}{2RR'}\biggr) e^{im(\phi-\phi')},
\end{equation}

\noindent where

\begin{equation}
\Psi\equiv
\frac{\bigl[(R^2-R'^2)^2+2(R^2+R'^2)(z-z')^2+(z-z')^4\bigl]^{\frac{1}{2}}} {2\pi\sqrt{RR'}}.
\end{equation}

\noindent This new formulation is amenable 
in analytic and computational physics applications on 
high-performance-computing architectures, since Laplacians can be 
readily computed on computational fluid dynamics meshes.  Higher order harmonic Green's
function expansions can be generated using this method.
Boundary values can now be computed effectively along an arbitrarily chosen 
z-axis $(m=0)$ and at values chosen to lie within the outer-most extent of
a chosen volumetric region $(m\geq 0)$.
Physical solution of the interior inhomogeneous problem can be obtained by 
proper boundary value treatment as described above.  Accelerations can be 
obtained from the potential by performing a precise numerical gradient.  In 
the case of cylindrical and spherical coordinates, the three dimensional 
Poisson solve can be further facilitated through the use of a discrete 
azimuthal Fourier transform.  This decouples the three dimensional Poisson 
problem into a set of decoupled 2D problems that can be solved with second 
order accurate finite differencing using either direct or iterative methods.

\section{Applications}

Using highly compact representations of infinite domain Green's 
functions, solutions to the linear partial differential equations of 
mathematical physics can be more easily obtained.  In this paper,
and in recent papers, we treat the Green's function for the Laplace
equation.  Here we have extended this result to the three dimensional biharmonic 
and higher harmonic equations.  Further variants are possible for the 
3D Helmholtz, wave, and diffusion equations.  In the future, we intend to 
investigate and more precisely describe these new Green's function 
expansions.  Many areas of theoretical physics will benefit greatly 
from precise numerical implementations of these compact expansions.

Azimuthal Fourier identities for potentials such as Coulomb
($|{\bf x}-{\bf x}'|^{-1}$)
or Yukawa
($e^{-k|{\bf x}-{\bf x}'|}|{\bf x}-{\bf x}'|^{-1}$)
lead to new classical 
and quantum energy theorems.  These allow for rapid and precise evaluation of 
the Coulomb or Yukawa direct and exchange interactions (Cohl et al. 2001).  
Classical Yukawa eigenfunctions are obtained through transcendental
function identification of the Lamb-Sommerfeld integral (see Magnus et al. 1966).
In quantum physics this is accomplished through the use of the azimuthal 
selection rule for the self-energies, namely for the direct and exchange 
Hamiltonian elements only the $m=0$ and $m=m_1-m_2$ terms survive respectively.  
The application of the selection rule allows for exact evaluation of the 
Hamiltonian matrix elements for two-electron interactions in atomic physics, 
molecular physics, condensed matter physics, physical chemistry, and biology.  
Two-electron interactions are critically important in obtaining opacities 
and correct equations of state in astrophysically dense atomic and 
molecular fluid media.  Magnetohydrodynamic problems can now be easily 
handled with Heine expansions for the Green's functions of potential theory.  
By expressing the equations of fluid dynamics of a compressible media in 
terms of a velocity potential and a vector Poisson equation, one may 
compute precise velocity boundary values in vortex and shock flow 
regions (Lamb 1932) by solving the appropriate Poisson problem.  
Radiation transport, and classical and quantum scattering, will be greatly 
facilitated through compact Poisson formulations of the Green's function 
for the 3D diffusion and Helmholtz equations.  Precise Coulomb and Yukawa
energies in the nuclear Hamiltonian allow for a higher degree of precision 
in obtaining nuclear structure.  

\section{Conclusion}

The ultimate resolution of these analytical investigations will be efficient 
algorithmic implementations of these new schemes.  We propose these methods 
to the three dimensional star community in hope that you may continue to 
enjoy significantly improved economical and precise boundary values for 
studies of analytical and numerical three dimensional stellar astrophysics.  

\acknowledgements

I would like to thank Ernie Kalnins, Department of Mathematics, University of 
Waikato, New Zealand, for pointing out the integration formula for Legendre 
functions.  I would also like to thank Volker Oberacker, Department of Physics
and Astronomy, Vanderbilt University, for pointing out the Green's function
for the axisymmetric biharmonic equation in terms of the complete elliptic 
integral of the second kind.  I especially wish to thank Dr. A.R.P. Rau,
Department of Physics and Astronomy, Louisiana State University, for many
helpful comments and suggestions.  I would also like to express my gratitude
to everyone I had the honor of meeting at the 3D stars workshop, the 
scientific organizing committee and great appreciation to the editors of 
this proceedings, Rob Cavallo, Stefan Keller, and Sylvain Turcotte. I would
also like to thank Richard Ward, David Dearborn, John Castor, John Bradley,
Kem Cook, Douglas Peters, Pete Eltgroth, and Peter Eggleton. 

This work was performed under the auspices of the U.S. Department of Energy,
National Nuclear Security Administration by the University of California,
Lawrence Livermore National Laboratory under contract No. W-7405-Eng-48.

\appendix
\section{Expressions for Unit Order Toroidal Functions}

The unit order negative one half degree Legendre function of the 
second kind can be expressed as

\begin{equation}
Q_{-\frac{1}{2}}^1(\zeta)=\frac{-1}{\sqrt{2(\zeta-1)}}E\biggl(\sqrt{\frac{2}{\zeta+1}}\biggr),
\end{equation}

\noindent where $E$ is the complete elliptic integral of the second kind.
The unit order, positive one half degree Legendre function can be expressed as

\begin{equation}
Q_{\frac{1}{2}}^1(\zeta)=
-\frac{\zeta}{\sqrt{2(\zeta-1)}}E\biggl(\sqrt{\frac{2}{\zeta+1}}\biggr)
+\sqrt{\frac{\zeta-1}{2}}K\biggl(\sqrt{\frac{2}{\zeta+1}}\biggr),
\end{equation}

\noindent where $K$ is the complete elliptic integral of the first kind.
We can express this same function in terms of the complete elliptic 
integral $D$

\begin{equation}
Q_{\frac{1}{2}}^1(\zeta)=
-\frac{1}{\sqrt{2(\zeta-1)}}E\biggl(\sqrt{\frac{2}{\zeta+1}}\biggr)
+\frac{\sqrt{2(\zeta-1)}}{\zeta+1}D\biggl(\sqrt{\frac{2}{\zeta+1}}\biggr),
\end{equation}

\noindent as well.  Higher degree, unit order toroidal functions of the second
kind can be easily derived using the following recurrence relation,

\begin{equation}
Q_{m+\frac{1}{2}}^1(\zeta)=\frac{4m\zeta}{2m-1}\ Q_{m-\frac{1}{2}}^1(\zeta) - 
\frac{2m+1}{2m-1}\ Q_{m-\frac{3}{2}}^1(\zeta).
\end{equation}

\vfill\eject

\begin{references}

\reference Abramowitz, M., \& Stegun, I. A. 1965, Handbook of Mathematical
Functions with Formulas, Graphs, and Mathematical Tables (New York: Dover)

\reference Cohl, H. S., \& Tohline, J. E. 1999, \apj, 527, 86 

\reference Cohl, H. S., Tohline, J. E., Rau, A. R. P., \& Srivastava, H. M. 2000,
Astron. Nachr., 321, 363

\reference Cohl, H. S., Rau, A. R. P., Tohline, J. E., Browne, D. A., 
Cazes, J. E., \& Barnes, E. I. 2001, \pra, 64, 52509

\reference Duffy, D. G. 2001, Green's Functions with Applications
(Boca Raton: CRC Press)

\reference Gradshteyn, I. S., \& Ryzhik, I. M. 1994, Table of Integrals,
Series, and Products (New York: Academic)

\reference Lamb, S. H. 1932, Hydrodynamics (Cambridge: Cambridge Univ. Press)

\reference Magnus, W., Oberhettinger, F., \& Soni, R. P. 1966, Formulas and 
Theorems for the Special Functions of Mathematical Physics (New York: Springer-Verlag)

\reference Miller, W., Jr. 1977, Symmetry and Separation of Variables
(New York: Academic)

\reference Morse, P., \& Feshbach, H. 1953, Methods of Theoretical Physics
(New York: McGraw-Hill)

\reference Vautherin, D. 1972, \prc, 7, 1, 296

\end{references}
\end{document}